\documentclass[preprint,showpacs,aps,prc]{revtex4}
\usepackage{amsmath}
\usepackage{graphics}
\input epsf

\def\be{\begin{equation}}
\def\ee{\end{equation}}
\def\ba{\begin{array}}
\def\ea{\end{array}}
\def\a{\alpha}
\def\q{Q_{\alpha}}
\def\hl{T_{1/2}^{}}

\def\hlex{T_{1/2}^{(expt)}}
\def\hlomp{T_{1/2}^{(OMP)}}
\def\hlanal{T_{1/2}^{(anal)}}
\def\hlexat{T_{1/2}^{(exat)}}

\def\lghlomp{log_{10}T_{1/2}^{(OMP)}}
\def\lghlanal{log_{10}T_{1/2}^{(anal)}}
\def\lghlexpt{log_{10}T_{1/2}^{(expt)}}

\begin{document}

\title
{The potential for $\a$ induced nuclear scattering, reaction and decay,
and a resonance-pole-decay model with exact explicit analytical solutions 
}

\author{Basudeb Sahu$^{1}$ and Swagatika Bhoi$^{2}$}
\affiliation{$^{1}$ Department of Physics, College of Engineering 
and Technology, Bhubaneswar-751003, India}

\affiliation{$^{2}$ School of Physics, Sambalpur University, Jyoti Vihar,  
Burla, 768019, India}

\begin{abstract}

The decay of $\a$ particle from a nucleus is viewed 
as a quantum resonance state of a two-body 
scattering process of the $\a$+daughter nucleus pair governed by a novel
nucleus-nucleus potential in squared Woods-Saxon form. By the application 
of the rigorous optical model (OM) potential scattering (S-matrix) theory 
the genuineness of the potential for the system 
is established by giving good explanation
of the elastic scattering and reaction cross sections data 
of the $\a$+nucleus pair.
From the pole position in the complex momentum (k) plane of the S-matrix 
defined above,  
the energy and width of the resonance state akin to 
the decaying state of emission of $\a$ particle are extracted
and from this width, 
the result of $\a$-decay half-life  
is derived to account for the experimental result of half-life in the cases
of large number of $\a$-emitters including heavy and super-heavy nuclei.
The S-matrix of the full OM calculation above is replaced by an
analytical function expressed in terms of exact Schr\"{o}dinger
solutions of a global potential that closely represents the 
Coulomb-nuclear interaction in the interior and the pure Coulomb wave
functions outside, and the resonant poles of this
S-matrix in the complex momentum plane are used to give satisfactory
results of decay half-lives of $\a$ coming out from varieties of
nuclei.\\

\noindent
PACS number(s): 23.60.+e, 21.10.Tg, 23.70.+j, 27.90.+b                        

\end{abstract}


\maketitle

\section{INTRODUCTION}
The process of decay of alpha ($\alpha$) particle from heavy and super-heavy
nuclei has been studied intensively in the past few years
\cite{moh06,xu05,den05,gam05,dup06,don05,dong05,chow06,sahu08,sahu10,sah16,
sah13}. 
In many papers
a simple two-body model was applied
\cite{sah16,sah13,gur87}
and in most papers a potential
was derived that was able to fit the measured $\a$-decay half-lives of the
alpha emitters. In recent time, such a potential for the $\a$+nucleus two-body
interaction is generated microscopically in the double folding ($t\rho \rho$
approximation) using explicitly the nuclear (both proton and neutron) 
densities 
\cite{gam03,gam05}.
However, most of the studies did not attempt to use these
potentials for the description of other experimental quantities such as,
for example, $\a$-scattering cross sections or reaction
(fusion) cross sections. Using the potential extracted from the fitting
of decay rate data, 
Devisov and Ikezoe \cite{den05} 
estimated the 
values of fusion cross section as a function of energy by treating
fusion process as a one-dimensional barrier passing mechanism.
Bhagwat and Gambhir \cite{bha08} 
have tried to account for the  measured results
of fusion cross sections in some cases of $\a$+nucleus systems
by the similar one-dimensional treatment of the fusion process and found
no success in explaining the data of fusion cross sections and the decay 
half-lives by using the potential obtained within the framework of
mean field theory. In both the studies stated above, the potential under
question has not been used or tested for the analysis of angular 
variation of the experimental values of differential scattering cross sections
at different incident energies. It is well known that the genuineness
of a nucleus-nucleus potential rests on the 
satisfactory explanation of the above
elastic scattering data in the optical model potential (OMP) analysis.
Through this analysis only, can one know
the exact height and radial position of the Coulomb-nuclear
potential barrier. Using a potential, without proper
verification of its barrier height and position, in the studies of 
other processes namely fusion and decay does not 
go well with the physical understanding of the processes.

In principle, the application of semi-classical model for tunneling is
not necessary for the calculation of $\a$-decay half-lives and the fusion
reaction cross section in the quantum mechanical two-body collision process
of $\a$-nucleus system. From the potential, using the rigorous
S-matrix (SM) theory of potential scattering, one can directly obtain the energy
and decay width from the poles of the SM in the lower half of the complex
momentum (k) plane close to the real axis 
\cite{sahu10}.
Further, this SM method can
be amalgamated within a code developed for calculating phase-shifts and cross
sections in the same $\a$+nucleus collision problem to explain the
elastic scattering and reaction (fusion) cross section data in a unified way
\cite{bsahu08,bsahu12}.
The motivation of this paper is to present a phenomenological
potential for the $\a$+nucleus system which is
consistent with the potential generated by relativistic mean field (RMF)
theory \cite{kum16}
and is suitable for 
simultaneous description of
three important events of $\a$ induced nuclear
reaction namely (i) elastic scattering, (ii) reaction, and (iii)
$\a$ emission by giving satisfactory explanation of the measured quantities:
elastic scattering cross section, reaction cross section and $\a$-decay 
half-life by the calculated results obtained using the SM theory of
potential scattering.

Further, the combined Coulomb-nuclear potential adopted above
is closely reproduced by a r-dependent potential expression. 
Using this potential form, we exactly solve the Schr\"{o}dinger equation 
and match them with the analytical
Coulomb wave functions outside and obtain
an expression for the S-matrix explicitly as a function of the incident
energy and the potential parameters. Then,
from the pole position in the complex momentum (k) plane of the S-matrix,
we extract the energy and width of the resonance state akin to 
the decaying state of emission of $\a$ particle from a nucleus.
With this simple model of potential scattering calculation we 
achieve good explanation
of the experimental results of $\a$-decay half-lives in the cases of 
several $\a$-emitters that include heavy and super-heavy nuclei.

In Sec. II, the details of the OMP
calculation 
and the derivation of the 
expression for the S-matrix of the exactly solvable potential are given.
Section III discusses the applications of the formulation to the
explanation of the experimental data of elastic scattering cross section,
reaction cross section and $\a$-decay half-life. In Sec. IV, we present the
summary and conclusion of the work.\\

\noindent
\section{Theoretical formulation}

The nuclear optical potential model is developed for the analysis
of the results of scattering and reaction cross sections obtained in the 
measurements of nucleus-nucleus collisions. In this quantum collision
theory, the following reduced
radial Schr\"{o}dinger equation
\be
\frac{d^2\phi(r)}{dr^2}+\frac{2\mu}{\hbar^2}(E-V(r))\phi(r)=0,
\ee
for a complex Coulomb-nuclear potential  
\be
V(r)=V_N(r)+V_C(r)+V_{\ell}(r),
\ee
the sum of the complex nuclear potential ($V_N(r)$), the
electrostatic potential ($V_C(r)$), and the centrifugal potential
($V_{\ell}(r)$) in the spatial region $0 < r \le R_{max}$, a distance where
the attractive nuclear potential becomes zero, is solved 
using the Runge-Kutta (RK) type of numerical integration or multistep
potential approximation
\cite{bs2008}
and the 
wave function ($\phi(r)$) and
its derivative ($d\phi(r)/dr$) at the radial position r=$R_{max}$ are obtained. 

In the outer region
$r > R_{max}$, the potential of the $\a$+nucleus interaction is only
Coulombic, V$_C(r)$, with the centrifugal term 
V$_{\ell}$=$\frac{\hbar^2}{2\mu}\frac{\ell(\ell+1)}{r^2}$
for different angular momentum partial wave $\ell$. Here, $\mu$ stands for the
reduced mass of the two-body system.
Using the exact Coulomb wave functions, i.e., F$_{\ell}(r)$ (regular)
and G$_{\ell}(r)$ (irregular) and their derivatives F$_{\ell}^{\prime}(r)$
and G$_{\ell}^{\prime}(r)$ in the outer region $r>R_{max}$
and the wave function $\phi(r)$ and its derivative 
$\frac{d\phi(r)}{dr}$
in the left side of $r=R_{max}$
and matching them at $r=R_{max}$, we get the expression for the partial
wave S-matrix denoted by S$_{\ell}$ as
\be
S_{\ell}=2i C_{\ell}+1,
\ee
where
\be
C_{\ell}=\frac{kF_{\ell}^{\prime}-F_{\ell}H}
{H(G_{\ell}+iF_{\ell})-k(G_{\ell}^{\prime}+iF_{\ell}^{\prime})},
\ee
\be
H=\frac{d\phi/dr}{\phi}\Big{\vert}_{r=R_{max}}.
\ee
with $k=\sqrt{\frac{2\mu}{\hbar^2}E}$ for the incident energy E.
In (4), prime ($^{\prime}$) denotes derivative with respect to $\rho$=kr.

We estimate the results of differential elastic scattering cross section
in ratio to Rutherford scattering as a function of scattering
angle ($\theta$) by using the S-matrix, S$_{\ell}$, given by (3)
in the following expression:
\be
\frac{d\sigma_{el}}{d\sigma_R}=\Big{\vert}\frac{i}{\eta}e^{-2i\sigma_0}
\{sin(\frac{\theta}{2})\}^{2(i\eta+1)}
\sum_0^{\infty}(2\ell+1)e^{2i\sigma_{\ell}}(S_{\ell}-1)P_{\ell}(cos\theta)
\Big{\vert}^{2}.
\ee
Here, the Sommerfeld parameter $\eta=\frac{\mu}{\hbar^2}\frac{Z_1Z_2e^2}{k}$
defined in terms of the wave number k, reduced mass $\mu$,
and proton numbers Z$_1$ and Z$_2$ of the two interacting nuclei with
the charge value e$^2$=1.4398 MeV fm. Further,
$\sigma_0$ stands for s-wave Coulomb phase-shift and is
expressed as 
\be
\sigma_0=\frac{\pi}{4}+\eta~log\eta -\eta-\frac{1}{12\eta}-\frac{1}{360\eta^3}
-\frac{1}{1260\eta^5}.
\ee
The Coulomb phase-shift, $\sigma_{\ell}$, for higher partial waves is 
evaluated using
\be
\sigma_{\ell}=\sigma_{\ell -1}+tan^{-1}\frac{\eta}{\ell}.
\ee
P$_{\ell}(cos\theta)$ stands for the Legendre polynomials.
For the total reaction cross section one can use the formula
\be
\sigma_R=\frac{\pi}{k^2}\sum_{\ell}(2\ell+1)(1-\mid S_{\ell}\mid^2).
\ee

In the optical model potential V(r) given by (2)
for the collision of two nuclei of mass
numbers A$_1$ and A$_2$ and proton number Z$_1$ and Z$_2$,
the complex nuclear potential
V$_N$(r)=V$_N^{R}$(r)+iV$_N^{I}$(r),
the sum of real part V$_N^R(r)$ and imaginary part V$_N^{I}(r)$.

We express
\be
V_N^R(r)=-V_0\frac{1+\delta~exp[-(r/R_v)^2]}{\left\{1+exp[(r-R_s)/2a_s]
\right\}^2},
\ee
\be
V_N^I(r)=-\frac{W_0}{\left\{1+exp[(r-R_I)/2a_I]\right\}^2}.
\ee
The radii R$_v$, R$_s$ and R$_I$ are expressed as 
R$_v=r_v(A_1^{1/3}+A_2^{1/3})$,
R$_s=r_s(A_1^{1/3}+A_2^{1/3})$, and
R$_I=r_I(A_1^{1/3}+A_2^{1/3})$, respectively in terms of distance parameters
r$_v$, r$_s$, and r$_I$ in fermi units.

The parameters a$_s$ and $a_I$ stand for the slope of the potentials
for the real and the imaginary parts, respectively. The depth parameters
V$_0 >0$ and W$_0 > 0$ and they are in energy (MeV) units. In the real
part V$_N^R(r)$ (10), there is a parameter $\delta$ which decides the depth
as V$_0(1+\delta)$ near the origin. The use of these potentials 
in squared Woods-Saxon form
has been found successful in the description of $\alpha+^{16}O$
elastic scattering and $\alpha$-cluster structure in $^{20}$Ne by 
Michel {\it et al.}, \cite{mich83}.
Thus, the real nuclear potential, V$_N^R(r)$ (10) is a 
five-parameter co-ordinate dependent expression with the 
adjustable parameters V$_0$, r$_v$, r$_s$, a$_s$, and $\delta$. 
The imaginary nuclear potential,
V$_N^{I}(r)$ (11) is a three-parameter 
formula with the parameters
W$_0$, r$_I$, and a$_I$ which are also adjustable.

The Coulomb potential, V$_C(r)$, based on homogeneous charge distributions
is expressed as
\be
V_C(r)=
\left\{
\begin{array}{cl}
\frac{Z_1Z_2e^2}{2R_C}(3-\frac{r^2}{R_C^2}), & {\rm if \; \; r \le R_C,}\\
\frac{Z_1Z_2e^2}{r},& {\rm if \; \; r > R_C,}
\end{array}
\right .
\ee
where radius parameter R$_C$=r$_C(A_1^{1/3}+A_2^{1/3})$
with r$_C\simeq$ 1.2 fm. 

Thus, the complete OMP, V(r) (2),
is specified by altogether nine parameters 
V$_0$, r$_s$, a$_s$, $\delta$, r$_v$,
r$_C$, W$_0$, r$_I$, and a$_I$.\\

\subsection{Poles of S-matrix for resonance and decay rate}

It may be mentioned here that the real nuclear potential V$_N^R(r)$ given 
by (10)
in combination with the Coulomb potential V$_C(r)$ (12) and the centrifugal
term V$_{\ell}(r)=\frac{\hbar^2}{2\mu}\frac{\ell(\ell+1)}{r^2}$,
generates a repulsive barrier in the outer region with a prominent pocket
in the inner side in each partial wave trajectory specified by $\ell=$0, 1,
2, 3, ....... The barrier along with the pocket found in a given $\ell$
gradually vanishes with the increase of $\ell$.
Such potentials with well defined pockets can generate significant resonances
in the reaction. These potential resonances are identified by the poles
of the S-matrix as defined below. Representing the S-matrix S$_{\ell}$ (3) by 
S$_{\ell}$(k)=$\frac{F(k)}{F(-k)}$,  
as a function of the wave number 
k, 
a zero at k=k$_r$-ik$_i$ of $F(-k)$ in the lower half of 
the complex k-plane gives rise to a pole in the S$_{\ell}$(k). When
k$_i<< $k$_r$, this corresponds to a resonance state with positive 
resonance energy E$_r$ and width $\Gamma_r$:
\be
E_r=(\frac{\hbar^2}{2\mu})(k_r^2-k_i^2),
\ee
\be
\Gamma_r=(\frac{\hbar^2}{2\mu})(4k_rk_i).
\ee
Thus, in the complex energy E plane S$_{\ell}$(k) has a pole at 
E=E$_r-i\Gamma_r/2$. The width $\Gamma_r$, expressed in energy unit, 
is related to decay constant $\lambda_d$, mean life T and half-life 
T$_{1/2}$ through the relation
\be
\frac{1}{\lambda_d}=T=T_{1/2}/0.693=\frac{\hbar}{\Gamma_r}.
\ee
\\

\subsection{Decay rate with exact explicit analytic solution}

Let us study the nature of the potential adopted above for the 
successful explanation of scattering, reaction and decay of an $\a$+nucleus
system. 
In Fig. 1, we plot the real nuclear part $V_N^R(r)$ (10) combined with
Coulomb potential $V_C(r)$ (12) for $\ell=0$ trajectory
with regard to the $\a$+daughter nucleus, 
$\a+^{208}$Pb, pair and show it by the dashed curve.
It is clearly seen that
there is a well defined pocket followed by a prominent barrier
with height V$_B$=20.27 MeV positioned at 
r=R$_B$=10.75 fm. This potential (dashed curve) is found close to the 
dotted curve which represents the potential calculated using energy 
density profiles of nucleons (proton and neutron) in RMF theory \cite{kum16}
except near the origin where the depth of the RMF potential is small.
As mentioned in Sec. II, this combined 
Coulomb-nuclear potential is responsible for generating resonances or 
quasi-molecular state that eventually decays. In order to match with the above
effective potential, we designate 
the following form  
\be
V_{eff}(r)=H_0\left\{\xi_1-(\xi_1-\xi_2)\rho(r)\right\},
\ee
where
 
$\rho(r)$=[cosh$^2(\frac{R_0-r}{d})]^{-1}$

\noindent 
is well known Eckart form factor and the strength parameter 
$H_0 >0$. This potential (16)
can be solved in the Schr\"{o}dinger equation
exactly. It has five parameters namely $H_0$, $\xi_1$, $\xi_2$,
$R_0$, and $d$.
With the values of the radial position of the barrier 
obtained by using global formula \cite{brog81}

R$_B$=$r_0(A_1^{1/3}+A_2^{1/3})$+2.72 fm 

\noindent
with $r_0=$1.07 fm, the height of the barrier 

V$_B$=$\frac{Z_1Z_2e^2}{R_B}(1-\frac{a_g}{R_B})$ 

\noindent
with $a_g=$0.63 fm
and setting R$_0$=R$_B$, H$_0\xi_2$=V$_B$, 
depth $H_0\xi_1$=$-$100 MeV, and
diffuseness d=9.63 fm,
the effective potential V$_{eff}(r)$ (16) is    
shown by a solid curve in Fig. 1 in the spatial region $0 < r < R_B$.
As we see, it closely matches, in region $0 < r < R_B$,
with the dashed curve that represents the
Coulomb+nuclear optical potential for $\ell$=0 described above.

For the potential expressed by Eq. (16), 
the s-wave radial Schr\"{o}dinger equation
can be written as 
\be
\frac{d^2u}{dr^2}+[\kappa^2-ik_0^2(1+i\xi)\rho(r)]u=0,
\ee
where

$\kappa^2=k^2-k_0^2\xi_1$,

$\xi=\xi_1-\xi_2$,

$k_0^2=\frac{2\mu}{\hbar^2}H_0$.

The exact solution of the above Eq. (17) is given as
\be
u(r)=AZ^{\frac{1}{2}\kappa d}F(a,b,c,Z)+BZ^{-\frac{1}{2}\kappa d}
F(a^{\prime},b^{\prime},c^{\prime},Z),
\ee
where Z=$[cosh^2\frac{R_0-r}{d}]^{-1}$ and
F(a,b,c,Z) is the hyper-geometric function.
The other terms are
\be
a=\frac{1}{2}(\lambda+i\kappa d),~~b=\frac{1}{2}(1-\lambda+i\kappa d),~~
c=1+i\kappa d,
\ee
\be
a^{\prime}=\frac{1}{2}(\lambda-i\kappa d),~~
b^{\prime}=\frac{1}{2}(1-\lambda-i\kappa d),~~
c^{\prime}=1-i\kappa d,
\ee
\be
\lambda=\frac{1}{2}-\frac{1}{2}[1-i(2k_0d)^2(i\xi)]^{1/2}.
\ee
Using the boundary condition

$u(r=0)=0$,

we get 
\be
Z(r=0)=Z_0=\frac{1}{cosh^2\frac{R_0}{d}},
\ee
and
\be
C=-\frac{B}{A}=Z_0^{i\kappa d}\frac{F(a,b,c,Z_0)}{F(a^{\prime},b^{\prime},
c^{\prime},Z_0)}.
\ee
For cosh$^2\frac{R_0}{d}>>1$, $Z_0<<1$,
\be
C\simeq exp(-2i\kappa [R_0-d~ln2]).
\ee
The logarithmic derivative of the wave function at r=$R_0$ is given by

\be
f(R_0)=\frac{du/dr}{u}\mid_{r=R_0}=
\frac{2}{d}\frac{\frac{\Gamma(c)}{\Gamma(a)\Gamma(b)}-C\frac{\Gamma(c^{\prime})}{\Gamma(a^{\prime})\Gamma(b^{\prime})}}
{\frac{\Gamma(c)}{\Gamma(c-a)\Gamma(c-b)}-C\frac{\Gamma(c^{\prime})}
{\Gamma(c^{\prime}-a^{\prime})\Gamma(c^{\prime}-b^{\prime})}}.
\ee
In the region $r>R_0=R_B$ where the potential is pure Coulombic, the
Coulomb wave functions (regular F$_0$ and irregular G$_0$)
and their derivatives (F$_0^{\prime}$ and G$_0^{\prime}$)
for $\ell=0$ case are expressed as
\cite{abr65}
\be
F_0=\frac{1}{2}\beta~ exp(\alpha),~~~
F_0^{\prime}=(\beta^{-2}+\frac{1}{8\eta}t^{-2}\beta^4)F_0,
\ee
\be
G_0=\frac{\beta}{exp(\alpha)},~~~
G_0^{\prime}=[-\beta^{-2}+\frac{1}{8\eta}t^{-2}\beta^4]G_0,
\ee
\be
t=\frac{\rho}{2\eta},~~~~~\beta=[\frac{t}{1-t}]^{\frac{1}{4}},
\ee
\be
\alpha=2\eta([t(1-t)]^{\frac{1}{2}}+arcsin~t^{\frac{1}{2}}-\frac{1}{2}\pi),
\ee
\be
k=\sqrt{\frac{2\mu}{\hbar^2}E},~~~~\rho=kR_0,~~~~\eta=\frac{\mu}{\hbar^2}
\frac{Z_1Z_2e^2}{k}.
\ee
By requirement of continuity at r=$R_0$, the wave functions and their
derivatives are matched at r=$R_0$ to obtain the scattering matrix
denoted by S(k) as
\be
S(k)=\frac{2ikF_0^{\prime}-2iF_0f(R_0)+f(R_0)[G_0+iF_0]-
k[G_0^{\prime}+iF_0^{\prime}]}
{f(R_0)[G_0+iF_0]-k[G_0^{\prime}+iF_0^{\prime}]},
\ee
where $f(R_0)$ is given by (25).
A pole of S(k) arising from the zero of the denominator of S(k) (31) in the 
lower half of  complex k-plane gives us the resonance energy equal to Q-value
and the decay half-life as described in Sec. II.\\

\section{Results and discussion}

In the application of the above formulation to the explanation of measured
data with regard to events namely scattering, reaction and $\a$-decay
in a $\a$+daughter nucleus system,
we select the $\a$+$^{208}$Pb reaction which has been subjected to extensive
experiments for the measurements of elastic scattering cross section,
reaction cross section, and rate of decay of $\a$ particle from the 
parent $^{212}$Po nucleus.

The values of the nine potential parameters we use for the total optical
model potential for this reaction are V$_0$= 22 MeV, r$_s$=1.27 fm, 
a$_s$=0.62 fm,
r$_v$=0.66 fm, $\delta$=3.5, r$_C$=1.2 fm, W$_0$=5 MeV, a$_I$=0.28 fm, and
r$_I$=1.2 fm for E$_{lab}$= 19 MeV,
r$_I$=1.3 fm for E$_{lab}$= 20 MeV,
r$_I$=1.37 fm for E$_{lab}$= 22 MeV where E$_{lab}$ stands for incident energy 
in the laboratory.

Using the S-matrix $S_{\ell}$ (3) in the expression (6),
we obtain the results of angular variation of differential elastic
scattering cross section $\sigma_{Sc}$ in ratio to
Rutherford scattering cross section $\sigma_{Ru}$ 
and compare them (full curves) with the
corresponding measured data (solid circles) obtained from Ref. \cite{bar1974}
in Fig. 2. 
It is found that the fitting of the data at three different
energies around the s-wave barrier height (=20.27 MeV) is quite good.
With this, the test for the authenticity of the nuclear optical potential
adopted in the present analysis is successful. Now the same potential
is to be tested for the explanation of reaction cross section and also
the result of $\a$-decay rate.

By using the expression (9), the total reaction cross section $\sigma_R$ as
a function of bombarding energy are obtained
and they are shown by a solid curve in Fig. 3 and compared with the 
corresponding experimental data represented by solid circles
\cite{bar1974} 
in this Fig. 3. It is clearly seen that the explanation of the data by
our calculated results (solid curve) is quite satisfactory
at different energies around the barrier. It may be pointed
out here that the results of $\sigma_{R}$ at low
($< 20 MeV$) incident
energies analyzed here are sometimes considered as fusion cross sections
\cite{bha08}.

Coming to the calculation of resonance energy and the decay 
width through the poles of S-matrix, we first find out the resonance energy
equal to the Q-value of $\a$-decay from the position of a peak in the 
variation of the result of $\sigma_R$ 
as a function of energy by using the same set of potential parameters
which is found successful in explaining the elastic and $\sigma_R$ 
data mentioned above.
In order to obtain the resonance energy exactly equal to the Q-value
we may need to marginally vary the depth or the diffuseness parameter
of the real nuclear potential which eventually does not affect the
results of elastic or reaction cross sections obtained earlier.
This resonance energy equal to the Q-value is used as a trial value of the 
real part of the pole position ($k_r^0$) of the S-matrix, $S_{\ell}$ (3).
The trial value of the imaginary part $(k_i^0$) of the pole position is 
taken to be a very small value ($\approx$ 0.001MeV) to begin with.
Starting with these trial values, Newton-Raphson iterative technique is 
used to obtain the zero of the Jost function of the Coulomb-nuclear
S-matrix, $S_{\ell}$ (3), which corresponds to the resonance or 
quasi-bound state pole. From this pole, using Eq. (13), the resonance
energy E$_r$ is obtained to represent the Q-value. The corresponding result
of width is obtained by using (14) and from this, using 
relation (15), the value of decay half-life
denoted by $\hlomp$ is obtained within the 
framework of optical model calculation. In the case of $\a$-emitter $^{212}$Po 
with Q-value $\q$=8.954 MeV, we find $\hlomp$= 2.89$\times 10^{-7}s$ 
which is very close to the experimental value of
half-life $\hlex$= 2.99$\times 10^{-7}s$.
For other $\a$+daughter nucleus pairs, we can use the same optical
model potential parameters used above in the analysis of the 
$\a$+$^{208}$Pb pair and estimate the $\a$-decay half-life 
$\hlomp$ of the parent nuclei.  Having fixed all other parameters
namely V$_0$=22 MeV, r$_s$=1.27 fm, a$_s$=0.62 fm, and r$_C$=1.2 fm, 
one has to marginally vary the 
value of the parameter r$_v$ around 0.66 fm and that of $\delta$ around
3 to generate the resonance energy exactly at the Q-value of a given pair.
This formulation can easily be applied to estimate the results of
decay half-lives of $\a$ particle emitted out with some angular momentum 
$\ell > 0$. For this, one has to simply generate the resonance at
the energy equal to the given Q-value of decay in the specified partial wave
trajectory $\ell$ by the variation of r$_v$ and $\delta$ outlined above.
We calculate the results of decay half-life in decimal logarithm,
($\lghlomp$), for several $\a$-emitters in the list of Polonium (Po) isotopes 
for $\ell$=0 state and compare them with
the corresponding experimental data denoted by 
$\lghlexpt$ in Table I. 
We find that our calculated results are close to the respective measured
data in most cases of $\a$+daughter nucleus pairs.

We now calculate $\hl$ of the $\a$+daughter system from the resonance
pole of the S-matrix, S(k) (31) derived by using exact wave functions
of Coulomb-nuclear interaction.
The same procedure adopted above in the OMP calculation is used here
to locate the pole of S(k) depicting the resonance at the 
energy equal to the Q-value of decay. In this case, the diffuseness parameter 
'd' of the potential (16) is varied to obtain the above situation of
resonance at the Q-value. From this pole of S(k) we derive decay $\hl$
by using formula (15) and denote the results by $\hlanal$ as it is based
on exact analytical solutions unlike those used in the derivation of
$\hlomp$ from the pole of S$_{\ell}$ (3) obtained within the OMP potential 
calculation.
In this potential model calculation, we obtain the result of $\hlanal$=
3.02$\times 10^{-7}s$
for the $\a$-decay of the $^{212}$Po nucleus which is very
close to the experimental result $\hlex$=2.99$\times 10^{-7}s$ with $\q$=8.954
MeV. Also, it is found that the result of $\hlanal$
is very close to the value of $\hlomp$=2.89$\times 10^{-7}$s. This 
closeness between our calculated results of $\hlanal$ and $\hlomp$
indicates that the
effective potential (16) with parameters r$_0$=1.07 fm, a$_g$=
0.63 fm and d=9.63 fm  
is a good approximation for the Coulomb-nuclear interaction potential
of the $\a$+$^{208}$Pb pair for the estimate of decay half-life
using exact solution of the potential for the S-matrix and its
resonance pole. 
As the potential (16) uses global formula
for its parameters for barrier position and height, one can use this in
the cases of other $\a$+daughter nucleus pairs with some variation of the 
diffuseness parameter d around 9 fm to generate the 
resonance at the energy equal to the Q-value of the of the given pair
and 
obtain the result of decay half-life from the pole of this resonance.
We calculate the results
($\hlanal$) for several $\a$-emitters in the list of Polonium (Po) isotopes 
for $\ell$=0 state and compare them with
the corresponding experimental data  
in Table I. 
We find that our calculated results of half-life in decimal logarithm,
($\lghlanal$),
having closely reproduced the results
($\lghlomp$) obtained in OMP calculation above
are found close to the respective measured
data $\lghlexpt$ in most cases of $\a$-emitters.

From the above analysis we learn  that for the
emission of $\a$ particle in $\ell$=0 situation,
instead of using poles from S$_{\ell}$ (3) which requires tedious
numerical calculations of wave functions of the full OMP,
it is okay to use the simple poles of S(k) (31)  
which is expressed analytically 
in terms of exact solution of a global potential and analytical Coulomb
wave functions, and
estimate the results of decay half-life,
$\hlanal$, for the explanation of experimental
decay rate in various $\a$+daughter nucleus pairs.
In Table II, we present the results of $\hlanal$
for several $\a$-emitters. On comparison with the corresponding
experimental data denoted by $\hlex$ in the same Table II, we find that our
results provide satisfactory explanation of the measured ones 
in most cases of the $\a$-emitters.

For emission of $\a$ particle with $\ell>0$, we have to use the
full optical potential in different trajectories specified by $\ell$s
and estimate the results, $\hlomp$, from the poles of S$_{\ell}$ (3) described
above. In Table III, we present our results of half-life, $\lghlomp$, in 
decimal logarithm for several
$\a$-emitters in different $\ell >$0 situations and compare them with the
corresponding experimental results denoted by $\lghlexpt$. We find that the
explanation of the measured data by our calculated ones is quite
satisfactory in most cases of decay. In few cases,
namely $^{159}_{73}X$, $^{214}_{91}X$, $^{229}_{91}X$, and $^{257}_{101}X$, 
the angular momenta assigned in the experimental results
are different from the $\ell$s we need to consider for proper fitting
of the $\q$-values and the corresponding half-lives. These $\ell$s decided
by our systematic calculation are noted within brackets () 
by the side of the measured $\ell$s for the above four nuclei.

Having obtained successes in the explanation of all the three $\a$ induced
nuclear collision events: elastic scattering, reaction and decay by the
use of the nuclear potential (10) in squared Woods-Saxon form, the following
few words are in order in favour of this potential.\\
\noindent
(i) The potential has a surface part defined by the parameters $V_0$,
$r_s$, and $a_s$ as in normal Woods-Saxon form and it takes care of 
proper description of the measured data of elastic scattering
which is a surface phenomenon.

\noindent
(ii) There is a volume part in this potential expression (10) governed
by the parameters r$_v$ and $\delta$ which controls the diffuseness of the
potential in the interior side. Interestingly, variation of these parameters
does not disturb much
the fitting of the elastic scattering cross section provided by the surface
part stated in point (i).
On the other hand, by selecting some values r$_v\sim$0.66 fm and $\delta\sim$
3, we, in combination with the repulsive Coulomb part, find an
effective potential which is slowly falling in nature towards the left hand
side of the Coulomb barrier. This bulging character of the Coulomb+nuclear
potential provides all the remarkable explanation of the experimental data
of $\a$-decay half-lives and also the reaction cross sections 
at energies near and below the barrier in large number of $\a$+daughter
nucleus collision events. From the successful application of this form
of nuclear potential we understand that the volume part 
deciding the diffuseness of the Coulomb barrier in the interior side
is the life-line for the explanation of the decay rate and reaction or 
fusion cross section in the collision of $\a$+nucleus system.

\noindent
(iii) This short of potential with less diffuseness in the interior side
of the Coulomb barrier potential is consistent
with the potential calculated using density profiles of 
nucleons in the RMF theory
\cite{kum16}.\\

\section{Summary and Conclusion}

Considering the process of decay of an $\a$ particle from a parent
nucleus as a two-body quantum collision of $\a$+daughter nucleus pair,
three events namely decay, elastic scattering and reaction (fusion) are
addressed in one platform within the framework of three-dimensional
optical model potential scattering (S-matrix) theory.
A novel expression for the nuclear potential in squared Woods-Saxon form
is adopted for the 
nucleus-nucleus collision.
Using the S-matrix of the complex nuclear plus electrostatic potentials,
the measured data of elastic scattering and reaction cross sections 
are explained to proof the genuineness of the potential. 
From the poles of the same S-matrix in the complex
momentum plane, 
we extract the energy and width of the resonance state akin to 
the decaying state of emission of $\a$ particle and from this width 
the result of decay half-life of the $\a$-emission 
is obtained to account for the experimental data of half-life in the cases
of large number of $\a$-emitters including heavy and super-heavy nuclei.

In this comprehensive analysis of three physical phenomena, we find that
the versatile form of nuclear potential adopted by us
in this paper, by virtue of its
surface part, explains data of elastic scattering cross section and by
the help of its volume part controlling the diffuseness of the potential in the
interior side, decides the results of reaction cross section
and decay rate yielding good explanation of respective
measured data. 

The sum of the above nuclear potential (real part) with the Coulomb
potential based on homogeneous distribution of charges for s-wave
is closely represented by an analytical expression as a function
of radial distance r which is solved exactly to express the S-matrix
in terms of the explicit analytical 
Schr\"{o}dinger solutions and Coulomb
wave functions. 
From the resonant poles of this well defined S-matrix of a global
soluble potential, the results of decay half-lives are obtained
to explain the corresponding experimental data in several
heavy as well as super-heavy $\a$ emitting  nuclei
giving rise to satisfactory explanation of the data.

In conclusion, we believe that
the emission of $\a$ particle from a radioactive
nucleus is governed by the fundamental principle
of quantal decay of charged particle from a resonance
state generated by a two-body ($\a$+daughter nucleus) potential that
describes the
elastic and
reaction cross sections of the $\a$+daughter nucleus collision.
And the width of the resonant pole of the S-matrix of the potential
yields the result of decay half-life.

Further, the S-matrix of the full optical model calculation 
involving tedious numerical computation for wave functions
can be replaced by a 
S-matrix which is expressed in terms of exact analytical 
solutions of a soluble potential that closely represents the real part of the 
potential describing elastic scattering data
and expressions of
Coulomb wave functions, and the resonant poles of this 
S-matrix in the complex momentum plane can be used to give satisfactory
results of $\a$-decay half-lives.\\

\centerline{ACKNOWLEDGMENTS}

We would like to thank Bharat Kumar, Ph.D. Scholar, 
Institute of Physics, Bhubaneswar, India, for supplying the 
results of the potential for the $\a$+$^{208}$Pb system using RMF theory.
We acknowledge the research facilities extended to us by the Institute of
Physics, Bhubaneswar, India.

\newpage
\begin{table}
{TABLE I.}
{Comparison of experimental results of
$\a$-decay half-life in decimal logarithm, $\lghlexpt$ (third column) 
with the calculated results
$\lghlomp$ (fourth column) obtained from the poles of S-matrix, S$_{\ell}$ (3),
and $\lghlanal$ (fifth column) from the poles of analytical S-matrix 
S(k) given by (31). 
In the sixth column, the values of diffuseness parameter
'd' used in the calculation of $\lghlanal$
are listed.
For the calculation of $\lghlomp$, the 
values of real optical potential parameters V$_0$=22 MeV, r$_s$=1.27 fm,
a$_s$=0.62 fm, and r$_C$=1.2 fm are kept same for
all nuclei, and the values of the parameters r$_v$ and $\delta$ are varied
around 0.66 fm and 3, respectively, to exactly reproduce the experimental 
$\q$ value of
a given alpha decaying isotope. Experimental data are obtained from Ref.
\cite{ren2012}. }\\
\begin{tabular}{cccccccccccccccccccccccc}
\hline
\hline
\multicolumn{1}{c}{Decay}&
\multicolumn{1}{c}{}&
\multicolumn{1}{c}{}&
\multicolumn{1}{c}{$Q_{\alpha}^{(expt)}$}&
\multicolumn{1}{c}{}&
\multicolumn{1}{c}{}&
\multicolumn{1}{c}{log$_{10}$T$_{1/2}^{(expt)}$}&
\multicolumn{1}{c}{}&
\multicolumn{1}{c}{}&
\multicolumn{1}{c}{log$_{10}$T$_{1/2}^{(OMP)}$}&
\multicolumn{1}{c}{}&
\multicolumn{1}{c}{}&
\multicolumn{1}{c}{log$_{10}$T$_{1/2}^{(anal)}$}&
\multicolumn{1}{c}{}&
\multicolumn{1}{c}{}&
\multicolumn{1}{c}{d}&\\
\multicolumn{1}{c}{}&
\multicolumn{1}{c}{}&
\multicolumn{1}{c}{}&
\multicolumn{1}{c}{(MeV)}&
\multicolumn{1}{c}{}&
\multicolumn{1}{c}{}&
\multicolumn{1}{c}{(s)}&
\multicolumn{1}{c}{}&
\multicolumn{1}{c}{}&
\multicolumn{1}{c}{(s)}&
\multicolumn{1}{c}{}&
\multicolumn{1}{c}{}&
\multicolumn{1}{c}{(s)}&
\multicolumn{1}{c}{}&
\multicolumn{1}{c}{}&
\multicolumn{1}{c}{(fm)}&\\
\hline
$^{218}$Po$\rightarrow$$^{214}$Pb &&& 6.115 &&&
2.27   &&& 2.36   &&& 2.28 &&& 7.9070 \\

$^{216}$Po$\rightarrow$$^{212}$Pb &&& 6.906 &&&
 $-$0.84  &&& $-$0.75   &&& $-$0.89 &&& 7.9915  \\

$^{214}$Po$\rightarrow$$^{210}$Pb &&& 7.833 &&&
$-$3.38   &&&$-$3.71    &&&$-$3.97  &&& 8.0996  \\

$^{212}$Po$\rightarrow$$^{208}$Pb &&& 8.954  &&&
$-$6.52   &&& $-$6.68   &&& $-$6.52 &&& 9.6357 \\

$^{210}$Po$\rightarrow$$^{206}$Pb &&&5.407  &&&
7.08   &&& 6.83   &&& 6.40 &&& 8.9432     \\

$^{208}$Po$\rightarrow$$^{204}$Pb &&&5.215  &&&
7.96   &&&7.96    &&& 7.47 &&& 8.8738     \\

$^{206}$Po$\rightarrow$$^{202}$Pb &&& 5.327  &&&
7.14   &&& 6.97   &&& 6.88 &&& 8.8554     \\

$^{204}$Po$\rightarrow$$^{200}$Pb &&& 5.485  &&&
6.28   &&& 6.14   &&& 6.05 &&& 8.8442     \\

$^{202}$Po$\rightarrow$$^{198}$Pb &&& 5.701 &&&
5.15   &&& 5.06   &&& 4.96 &&& 8.8425     \\

$^{200}$Po$\rightarrow$$^{196}$Pb &&& 5.981 &&&
3.74   &&& 3.74   &&& 3.64 &&& 8.8511     \\

$^{198}$Po$\rightarrow$$^{194}$Pb &&& 6.309 &&&
2.27   &&& 2.30   &&& 2.19 &&& 8.8677 \\

$^{196}$Po$\rightarrow$$^{192}$Pb &&& 6.657  &&&
0.77   &&& 0.90   &&& 0.79 &&& 8.8765  \\

$^{194}$Po$\rightarrow$$^{190}$Pb &&& 6.987 &&&
$-$0.41   &&& $-$0.31   &&& $-$0.43 &&& 8.9045 \\

$^{190}$Po$\rightarrow$$^{186}$Pb &&& 7.693 &&&
$-$2.61   &&& $-$2.65   &&& $-$2.77 &&& 8.9461 \\
\hline
\hline
\end{tabular}
\end{table}
                           
\newpage
\begin{table}
{TABLE II.}
{Comparison of the experimental $\a$-decay half-lives 
with the calculated
ones for nuclei with the neutron number
N $>$ 126. The first and second columns denote the elemental symbol and 
the mass number of the parent nucleus. The third and fourth columns are,
respectively, the experimental decay energies ($\q$ values) and half-lives
($\hlex$) of $\a$ decay obtained from
Ref. \cite{ni2009}. The half-lives, $\hlanal$, calculated from the 
poles of analytical S-matrix S(k) given by (31) are presented in the 
fifth column. In the sixth column, the values of diffuseness parameter
'd' used in the calculation are listed.$~~~~~~~~~~~~~~~~~~~~~~~~~~~~~~~~~~~~~~~~~~~~~~~$}\\
 
\begin{tabular}{cccccccccccccccccccccccc}
\hline
\hline
\multicolumn{1}{c}{Elt.}&
\multicolumn{1}{c}{}&
\multicolumn{1}{c}{}&
\multicolumn{1}{c}{}&
\multicolumn{1}{c}{A}&
\multicolumn{1}{c}{}&
\multicolumn{1}{c}{}&
\multicolumn{1}{c}{}&
\multicolumn{1}{c}{$\q$(MeV)}&
\multicolumn{1}{c}{}&
\multicolumn{1}{c}{}&
\multicolumn{1}{c}{}&
\multicolumn{1}{c}{$\hlex $(s)}&
\multicolumn{1}{c}{}&
\multicolumn{1}{c}{}&
\multicolumn{1}{c}{}&
\multicolumn{1}{c}{$\hlanal $(s)}&
\multicolumn{1}{c}{}&
\multicolumn{1}{c}{}&
\multicolumn{1}{c}{}&
\multicolumn{1}{c}{d(fm)}&\\
\hline
\hline
Pb&&&&210 &&&&3.792 &&&&3.69$\times10^{16}$&&&&3.00$\times10^{16}$&&&&8.73\\
Po&&&&212 &&&&8.954 &&&&2.99$\times10^{-7}$ &&&&3.02 $\times10^{-7}$&&&&9.63\\
 &&&&214 &&&&7.833 &&&&1.64$\times10^{-4}$ &&&&1.02 $\times10^{-4}$&&&&8.10\\
 &&&&216 &&&&6.906 &&&&1.45$\times10^{-1}$ &&&&1.24 $\times10^{-1}$&&&&7.99\\
 &&&&218&&&&6.115 &&&&1.86$\times10^{2}$ &&&&1.91 $\times10^{2}$&&&&7.90\\
Rn&&&&214 &&&&9.208 &&&&2.70 $\times10^{-7}$ &&&&1.22 $\times10^{-7}$&&&&8.26\\
 &&&&216  &&&&8.200 &&&&4.50 $\times10^{-5}$ &&&&5.10 $\times10^{-5}$&&&&8.14 \\
 &&&&218 &&&&7.263 &&&&3.5 $\times10^{-2}$ &&&&4.68 $\times10^{-2}$&&&&8.03 \\
 &&&&220 &&&&6.405 &&&&5.56 $\times10^{1}$ &&&&9.1 $\times10^{1}$&&&&7.93 \\
 &&&&222  &&&&5.590 &&&&3.31 $\times10^{5}$ &&&&6.1 $\times10^{5}$&&&&7.85 \\
Ra&&&&216 &&&&9.526 &&&&1.82 $\times10^{-7}$ &&&&1.05 $\times10^{-7}$&&&&8.29\\
 &&&&218 &&&&8.546 &&&&2.56 $\times10^{-5}$ &&&&3.10 $\times10^{-5}$&&&&8.17 \\
 &&&&220  &&&&7.592 &&&&1.81 $\times10^{-2}$ &&&&2.36 $\times10^{-2}$&&&&8.06\\
 &&&&222  &&&&6.679 &&&&3.92 $\times10^{1}$ &&&&5.37 $\times 10^{1}$&&&&7.96\\
 &&&&224 &&&&5.789 &&&&3.33 $\times10^{5}$ &&&&5.96 $\times 10^{5}$&&&&7.86\\
&&&&226 &&&&4.871 &&&&5.35 $\times10^{10}$ &&&&12.56 $\times 10^{10}$&&&&7.76\\
Th&&&&218 &&&&9.849 &&&&1.09 $\times10^{-7}$ &&&&0.89 $\times 10^{-7}$&&&&8.33\\
&&&&220 &&&&8.953 &&&&9.70 $\times10^{-6}$ &&&&1.33 $\times10^{-5}$&&&&8.22\\
&&&&222 &&&&8.127 &&&&2.05 $\times10^{-3}$ &&&&2.80 $\times10^{-3}$&&&&8.13 \\
&&&&224 &&&&7.298 &&&&1.33 $\times10^{0}$ &&&&1.61 $\times10^{0}$&&&&8.03 \\
&&&&226 &&&&6.451 &&&&2.46 $\times10^{3}$ &&&&3.94 $\times10^{3}$&&&&7.94 \\
&&&&228 &&&&5.520 &&&&8.49 $\times10^{7}$ &&&&1.65 $\times10^{8}$&&&&7.84 \\
\hline
\hline
\end{tabular}
\end{table}

\newpage
\begin{table}
{~~~~TABLE II.}{~~~~~~~~~~~~~~~~~~~~~~~~~~(continued).}
\begin{tabular}{cccccccccccccccccccccccc}
\hline
\hline
\multicolumn{1}{c}{Elt.}&
\multicolumn{1}{c}{}&
\multicolumn{1}{c}{}&
\multicolumn{1}{c}{}&
\multicolumn{1}{c}{A}&
\multicolumn{1}{c}{}&
\multicolumn{1}{c}{}&
\multicolumn{1}{c}{}&
\multicolumn{1}{c}{Q(MeV)}&
\multicolumn{1}{c}{}&
\multicolumn{1}{c}{}&
\multicolumn{1}{c}{}&
\multicolumn{1}{c}{$\hlex $(s)}&
\multicolumn{1}{c}{}&
\multicolumn{1}{c}{}&
\multicolumn{1}{c}{}&
\multicolumn{1}{c}{$\hlexat $(s)}&
\multicolumn{1}{c}{}&
\multicolumn{1}{c}{}&
\multicolumn{1}{c}{}&
\multicolumn{1}{c}{d(fm)}&\\
\hline
\hline
&&&&230 &&&&4.770 &&&&3.12 $\times10^{12}$ &&&&7.82 $\times10^{12}$&&&&7.77 \\
&&&&232 &&&&4.082 &&&&5.69 $\times10^{17}$ &&&&2.21 $\times10^{18}$&&&&7.71 \\
U&&&&222 &&&&9.500 &&&&1.40 $\times10^{-6}$ &&&&2.62 $\times10^{-6}$&&&&8.29 \\
&&&&224 &&&&8.620 &&&&9.40 $\times10^{-4}$ &&&&5.66 $\times10^{-4}$&&&&8.18 \\
&&&&226 &&&&7.701 &&&&2.69 $\times10^{-1}$ &&&&4.14 $\times10^{-1}$&&&&8.08 \\
&&&&228 &&&&6.803 &&&&5.75 $\times10^{2}$ &&&&9.78 $\times10^{2}$&&&&7.98 \\
&&&&230 &&&&5.993 &&&&2.67 $\times10^{6}$ &&&&5.08 $\times10^{6}$&&&&7.90 \\
&&&&232 &&&&5.414 &&&&3.20 $\times10^{9}$ &&&&7.32 $\times10^{9}$&&&&7.85 \\
&&&&234 &&&&4.858 &&&&1.09 $\times10^{13}$ &&&&2.52 $\times10^{13}$&&&&7.80 \\
&&&&236 &&&&4.673 &&&&1.00 $\times10^{15}$ &&&&3.08 $\times10^{15}$&&&&7.79 \\
&&&&238 &&&&4.270 &&&&1.78 $\times10^{17}$ &&&&8.82 $\times10^{17}$&&&&7.78 \\
Pu&&&&232 &&&&6.716 &&&&1.71 $\times10^{4}$ &&&&1.77 $\times10^{4}$&&&&7.99 \\
&&&&234 &&&&6.310 &&&&7.73 $\times10^{5}$ &&&&12.31 $\times10^{5}$&&&&7.96 \\
&&&&236 &&&&5.867 &&&&1.30 $\times10^{8}$ &&&&2.11 $\times10^{8}$&&&&7.93 \\
&&&&238 &&&&5.593 &&&&3.90 $\times10^{9}$ &&&&6.62 $\times10^{9}$&&&&7.92 \\
&&&&240 &&&&5.256 &&&&2.84 $\times10^{11}$ &&&&6.82 $\times10^{11}$&&&&7.91 \\
&&&&242 &&&&4.985 &&&&1.52 $\times10^{13}$ &&&&3.95 $\times10^{13}$&&&&7.90 \\
&&&&244 &&&&4.666 &&&&3.17 $\times10^{15}$ &&&&8.11 $\times10^{15}$&&&&7.88 \\
Cm&&&&240 &&&&6.398 &&&&3.30 $\times10^{6}$ &&&&3.75 $\times10^{6}$&&&&8.02 \\
&&&&242 &&&&6.216 &&&&1.90 $\times10^{7}$ &&&&2.82 $\times10^{7}$&&&&8.03 \\
&&&&244 &&&&5.902 &&&&7.48 $\times10^{8}$ &&&&1.16 $\times10^{9}$&&&&8.01 \\
&&&&246 &&&&5.475 &&&&1.82 $\times10^{11}$ &&&&3.20 $\times10^{11}$&&&&7.98 \\
&&&&248 &&&&5.162 &&&&1.43 $\times10^{13}$ &&&&3.03 $\times10^{13}$&&&&7.97 \\
Cf&&&&240 &&&&7.719 &&&&9.09 $\times10^{1}$ &&&&8.77 $\times10^{1}$&&&&8.17 \\
&&&&242 &&&&7.517 &&&&2.62 $\times10^{2}$ &&&&4.77 $\times10^{2}$&&&&8.17 \\
&&&&244 &&&&7.329 &&&&1.55 $\times10^{3}$ &&&&2.45 $\times10^{3}$&&&&8.17 \\
&&&&246 &&&&6.862 &&&&1.62 $\times10^{6}$ &&&&2.12 $\times10^{5}$&&&&8.14 \\
&&&&248 &&&&6.361 &&&&3.54 $\times10^{7}$ &&&&4.37 $\times10^{7}$&&&&8.10 \\
\hline
\hline
\end{tabular}
\end{table}
                        
\newpage
\begin{table}
{~~~~TABLE II.}{~~~~~~~~~~~~~~~~~~~~~~~~~~(continued).}
\begin{tabular}{cccccccccccccccccccccccc}
\hline
\hline
\multicolumn{1}{c}{Elt.}&
\multicolumn{1}{c}{}&
\multicolumn{1}{c}{}&
\multicolumn{1}{c}{}&
\multicolumn{1}{c}{A}&
\multicolumn{1}{c}{}&
\multicolumn{1}{c}{}&
\multicolumn{1}{c}{}&
\multicolumn{1}{c}{$\q$(MeV)}&
\multicolumn{1}{c}{}&
\multicolumn{1}{c}{}&
\multicolumn{1}{c}{}&
\multicolumn{1}{c}{$\hlex $(s)}&
\multicolumn{1}{c}{}&
\multicolumn{1}{c}{}&
\multicolumn{1}{c}{}&
\multicolumn{1}{c}{$\hlexat $(s)}&
\multicolumn{1}{c}{}&
\multicolumn{1}{c}{}&
\multicolumn{1}{c}{}&
\multicolumn{1}{c}{d(fm)}&\\
\hline
\hline
&&&&250 &&&&6.128 &&&&4.88 $\times10^{8}$ &&&&6.48 $\times10^{8}$&&&&8.09 \\
&&&&252 &&&&6.217 &&&&1.02 $\times10^{8}$ &&&&2.10 $\times10^{8}$&&&&8.14 \\
&&&&254 &&&&5.927 &&&&2.04 $\times10^{9}$ &&&&7.07 $\times10^{9}$&&&&8.12 \\
Fm&&&&246 &&&&8.374 &&&&1.55 $\times10^{0}$ &&&&2.69 $\times10^{0}$&&&&8.31 \\
&&&&248 &&&&8.002 &&&&4.56 $\times10^{1}$ &&&&4.67 $\times10^{1}$&&&&8.29 \\
&&&&250 &&&&7.557 &&&&2.28 $\times10^{3}$ &&&&2.08 $\times10^{3}$&&&&8.25 \\
&&&&252 &&&&7.153 &&&&1.09 $\times10^{5}$ &&&&8.30 $\times10^{4}$&&&&8.22 \\
&&&&254 &&&&7.308 &&&&1.37 $\times10^{4}$ &&&&1.78 $\times10^{4}$&&&&8.28 \\
&&&&256 &&&&7.027 &&&&1.35 $\times10^{5}$ &&&&2.78 $\times10^{5}$&&&&8.27 \\
No&&&&252 &&&&8.550 &&&&4.18 $\times10^{0}$ &&&&3.96 $\times10^{0}$&&&&8.38 \\
&&&&254 &&&&8.226 &&&&7.14 $\times10^{1}$ &&&&4.58 $\times10^{1}$&&&&8.37 \\
&&&&256 &&&&8.581 &&&&3.64 $\times10^{0}$ &&&&2.89 $\times10^{0}$&&&&8.45 \\
Rf&&&&256 &&&&8.930 &&&&2.02 $\times10^{0}$ &&&&1.32 $\times10^{0}$&&&&8.46 \\
&&&&258 &&&&9.250 &&&&9.23 $\times10^{-2}$ &&&&1.36 $\times10^{-1}$&&&&8.54 \\
\hline
\hline
\end{tabular}
\end{table}

\newpage
\begin{table}
{TABLE III.}
{Comparison of experimental values in decimal logarithm
log$_{10}T^{(expt)}_{1/2}$
of half-life of $\a$-decay and  corresponding results
of present calculation
log$_{10}T^{(OMP)}_{1/2}$
obtained from the poles of S$_{\ell}$ (3) in decimal logarithm.
The experimental $\q$ values, half-lives and $\l$ values are obtained from
Ref. \cite{roy2010}.}
\begin{tabular}{cccccccccccccccccccccccc}
\hline
\hline
\multicolumn{1}{c}{$^{A}_{Z}$}&
\multicolumn{1}{c}{}&
\multicolumn{1}{c}{$\q$(MeV)}&
\multicolumn{1}{c}{}&
\multicolumn{1}{c}{$\l$}&
\multicolumn{1}{c}{}&
\multicolumn{1}{c}{log$_{10}T^{(expt)}_{1/2}(s)$}&
\multicolumn{1}{c}{}&
\multicolumn{1}{c}{log$_{10}T^{(OMP)}_{1/2}(s)$}&
\multicolumn{1}{c}{}&
\multicolumn{1}{c}{$^{A}_{Z}$}&
\multicolumn{1}{c}{}&
\multicolumn{1}{c}{$\q$(MeV)}&
\multicolumn{1}{c}{}&
\multicolumn{1}{c}{$\l$}&
\multicolumn{1}{c}{}&
\multicolumn{1}{c}{log$_{10}T^{(expt)}_{1/2}(s)$}&
\multicolumn{1}{c}{}&
\multicolumn{1}{c}{log$_{10}T^{(OMP)}_{1/2}(s)$}&\\
\hline
\hline
  $^{112}_{   53}$&&   2.990 &&  4 &&  5.45 &&  5.54&&
  $^{149}_{   65}$&&   4.077&&   2 &&  4.97 &&  4.96\\
  $^{151}_{   65}$&&   3.496 &&  2 &&  8.82 &&  8.80&&
  $^{159}_{   73}$&&   5.681 &&  5(0) &&  0.11 && 0.12\\
  $^{162}_{   73}$&&   5.010 &&  1 &&  3.68 &&  3.62&&
  $^{175}_{   77}$&&   5.400 &&  2 &&  3.02 &&  3.38\\
  $^{181}_{   79}$&&   5.751&&   2 &&  3.39 &&  3.34&&
  $^{191}_{   83}$&&   6.778&&   5 &&  2.85 && 2.96\\
  $^{193}_{   83}$&&   6.304&&   5 &&  4.50 &&  4.22&&
  $^{195}_{   83}$&&   5.832&&   5 &&  6.79 && 6.34\\
  $^{210}_{   85}$&&   5.631&&   2 &&  7.73 &&  7.34&&
  $^{210}_{   87}$&&   6.650&&   2 &&  2.43 &&  2.69\\
  $^{212}_{   83}$&&   6.207&&   5 &&  4.57 &&  4.18&&
  $^{212}_{   85}$&&   7.824&&   5 &&-0.42&&  -0.42\\
  $^{212}_{   87}$&&   6.529&&   2 &&  4.10 &&  4.10&&
  $^{213}_{   83}$&&   5.982&&   5 &&  5.15 &&  5.18&&\\
  $^{214}_{   83}$&&   5.621&&   5 &&  7.16 &&  7.20&&
  $^{214}_{   87}$&&   8.589&&   5 && -2.27 && -2.20&&\\
  $^{214}_{   89}$&&   7.350&&   2 &&  1.23 &&  1.35&&
  $^{214}_{   91}$&&   8.430&&   4(1) && -2.10 && -2.32&&\\
  $^{216}_{   89}$&&   9.235&&   5 && -3.31 && -3.30&&
  $^{220}_{   87}$&&   6.801&&   1 &&  1.62 &&  1.82&&\\
  $^{221}_{   87}$&&   6.457 &&  2 &&  2.55 &&  2.73&&
  $^{223}_{   89}$&&   6.783 &&  2 &&  2.60 &&  2.51&&\\
  $^{224}_{   89}$&&   6.327 &&  1 &&  5.73 &&  5.31&&
  $^{225}_{   89}$&&   5.935&&   2 &&  6.23 &&  6.34&&\\
  $^{225}_{   91}$&&   7.390&&   2 &&  0.39 && 0.47 &&
  $^{226}_{   89}$&&   5.537&&   2 &&  9.25 &&  9.42&&\\
  $^{228}_{   91}$&&   6.264&&   3 &&  7.60 && 7.23&&
  $^{229}_{   91}$&&   5.835 &&  1(3) && 10.03 &&  10.02&&\\
  $^{230}_{   91}$&&   5.439 &&  2 && 11.31 &&  10.95&&
  $^{235}_{   93}$&&   5.194 &&  1 && 13.94 && 13.62&&\\
  $^{235}_{   95}$&&   6.610 &&  1 &&  5.17 && 5.35 &&
  $^{237}_{   93}$&&   4.958 &&  1 && 16.19 && 15.75&&\\
  $^{239}_{   95}$&&   5.922 &&  1 && 11.11 && 10.92&&
  $^{241}_{   95}$&&   5.638 &&  1 && 12.60 && 12.50&&\\
  $^{243}_{   95}$&&   5.439 &&  1 && 14.16 && 13.67&&
  $^{245}_{   97}$&&   6.455 &&  2 &&  9.37 && 9.79&&\\
  $^{245}_{   99}$&&   7.909 &&  3 &&  3.52 && 3.37 &&
  $^{249}_{   97}$&&   5.525 &&  2 && 13.61 && 13.55&&\\
  $^{252}_{   99}$&&   6.790 &&  1 &&  7.83 && 7.85 &&
  $^{257}_{  101}$&&   7.558 &&  1(4) &&  7.57 && 7.83&&\\
\hline
\hline
\end{tabular}
\end{table}

\newpage
\begin{figure}
\begin{center}
\leavevmode
\hbox{\epsfxsize=6.5in
\epsfysize=6.50in
\centerline {\epsffile {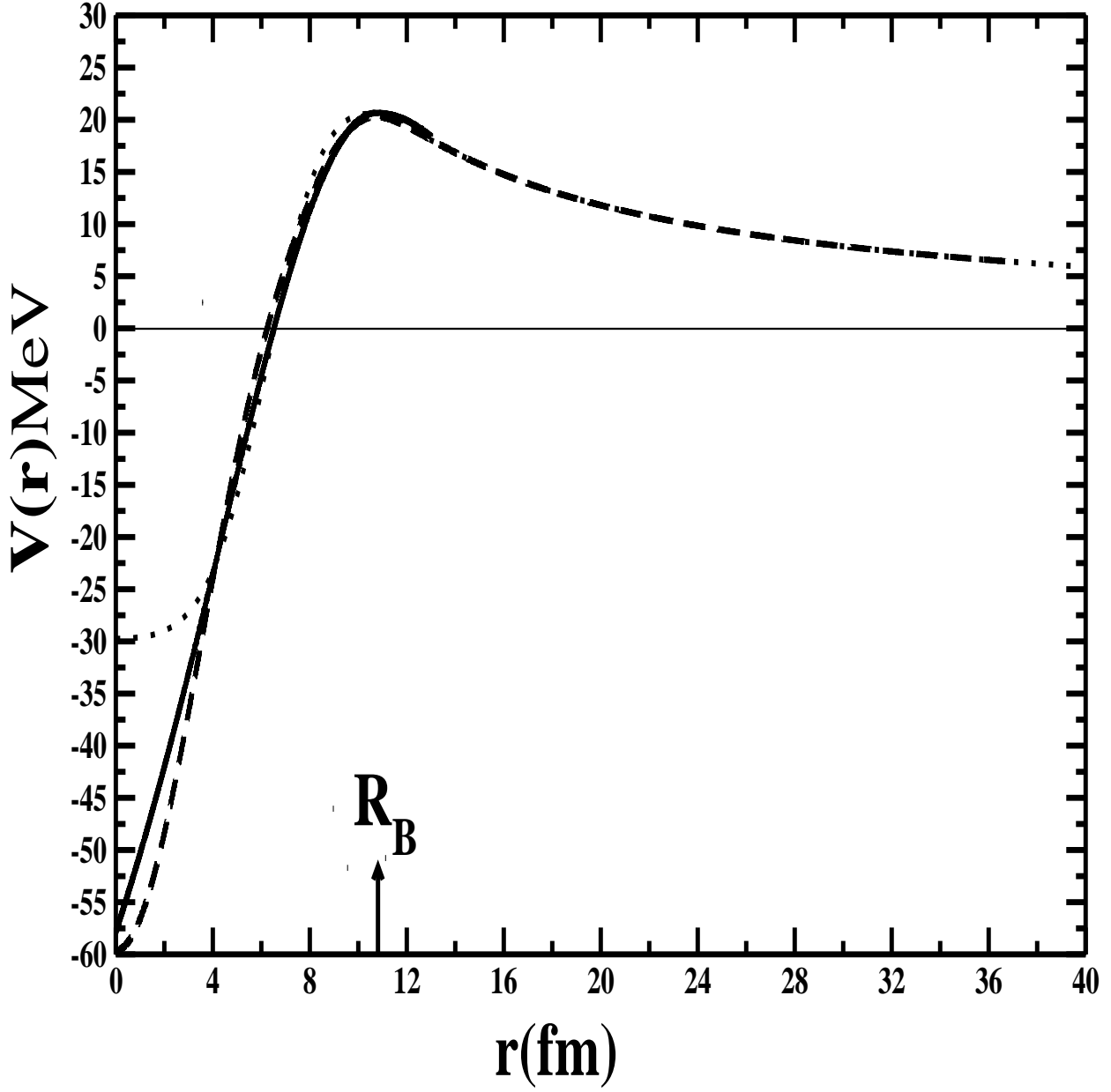}}}
\end{center}
\vspace*{0.2 true in}
\caption{Plot of potential V(r) as a function of radial distance for
the $\a$+$^{208}$Pb system. The dashed curve represents the sum of 
real nuclear potential expressed by (10) with parameters V$_0=$22 MeV,
a$_s$=0.62 fm, r$_s$=1.27 fm, r$_v$=0.66 fm, $\delta$=3.5 and the Coulomb
potential given by (12) with r$_C$=1.2 fm. The solid curve represents the
potential expressed by (16) in the text with parameters r$_0$=1.07 fm,
a$_g$=0.63 fm, H$_0$=1 MeV, $\xi_1=-$100, and $\xi_2$=V$_B$=20.27. The arrow 
indicates barrier position R$_B$=10.75 fm.
The dotted curve represents the potential calculated using  
RMF theory \cite{kum16}.
}
\end{figure}

\newpage
\begin{figure}
\begin{center}
\leavevmode
\hbox{\epsfxsize=6.5in
\epsfysize=5.0in
\centerline {\epsffile {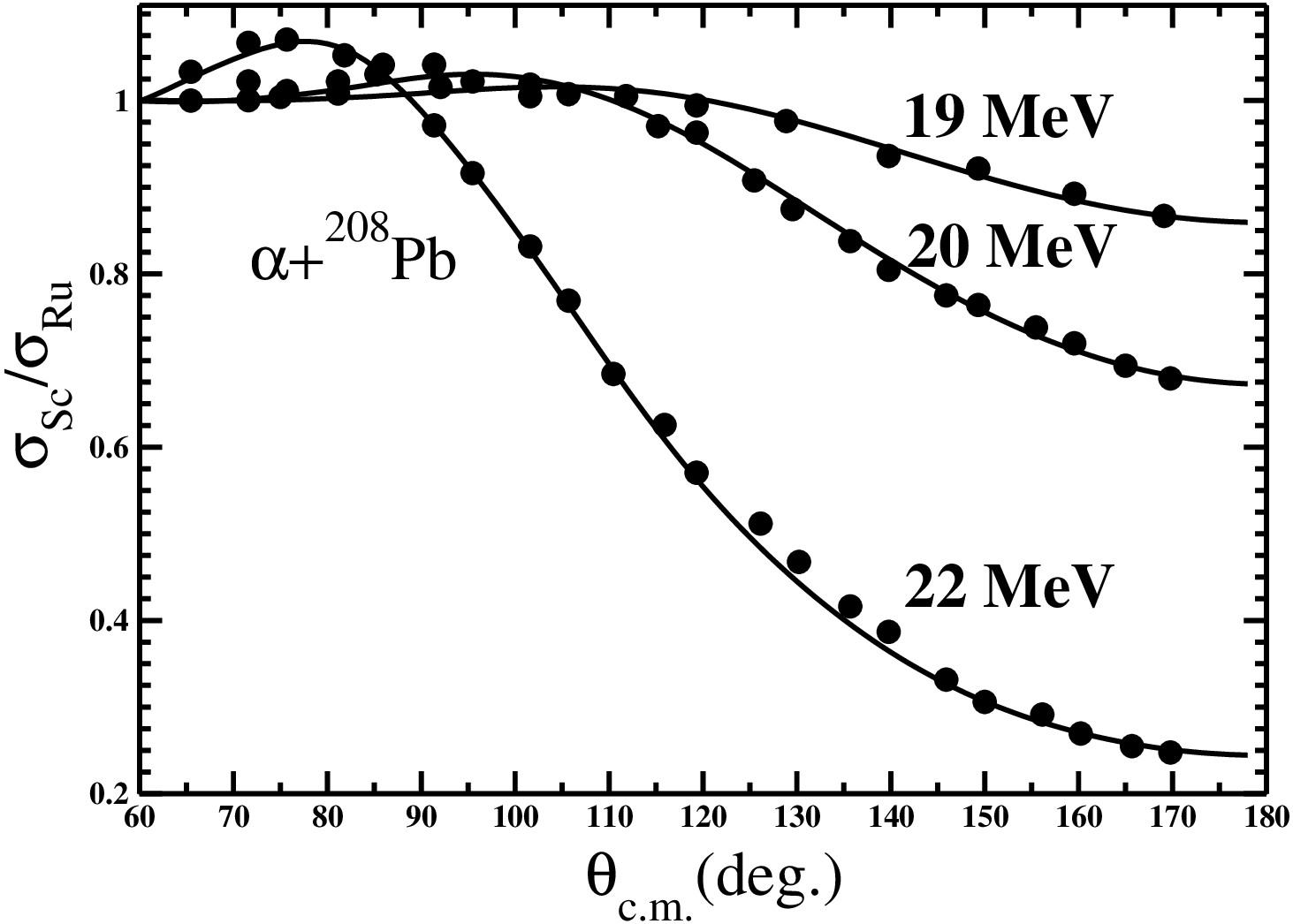}}}
\end{center}
\vspace*{0.2 true in}
\caption{ Angular variation of elastic differential scattering cross
section $\sigma_{Sc}$, relative to Rutherford $\sigma_{Ru}$, of $\a$
particles scattered by $^{208}$Pb at 19, 20 and 22 Mev laboratory energies.
The solid curves represent the results of present
optical model calculations using the parameters:
$V_0$= 22 MeV, $r_s$=1.27 fm, $a_s$=0.62 fm,
$r_v$=0.66 fm, $\delta$=3.5, $r_C$=1.2 fm, $W_0$=5 MeV, $a_I$=0.28 fm, 
$r_I$=1.2 fm for 19 MeV,
$r_I$=1.3 fm for 20 MeV, and
$r_I$=1.37 fm for 22 MeV.
The experimental data shown by solid circles are obtained from \cite{bar1974}.
}
\end{figure}

\newpage
\begin{figure}
\begin{center}
\leavevmode
\hbox{\epsfxsize=6.5in
\epsfysize=5.50in
\centerline {\epsffile {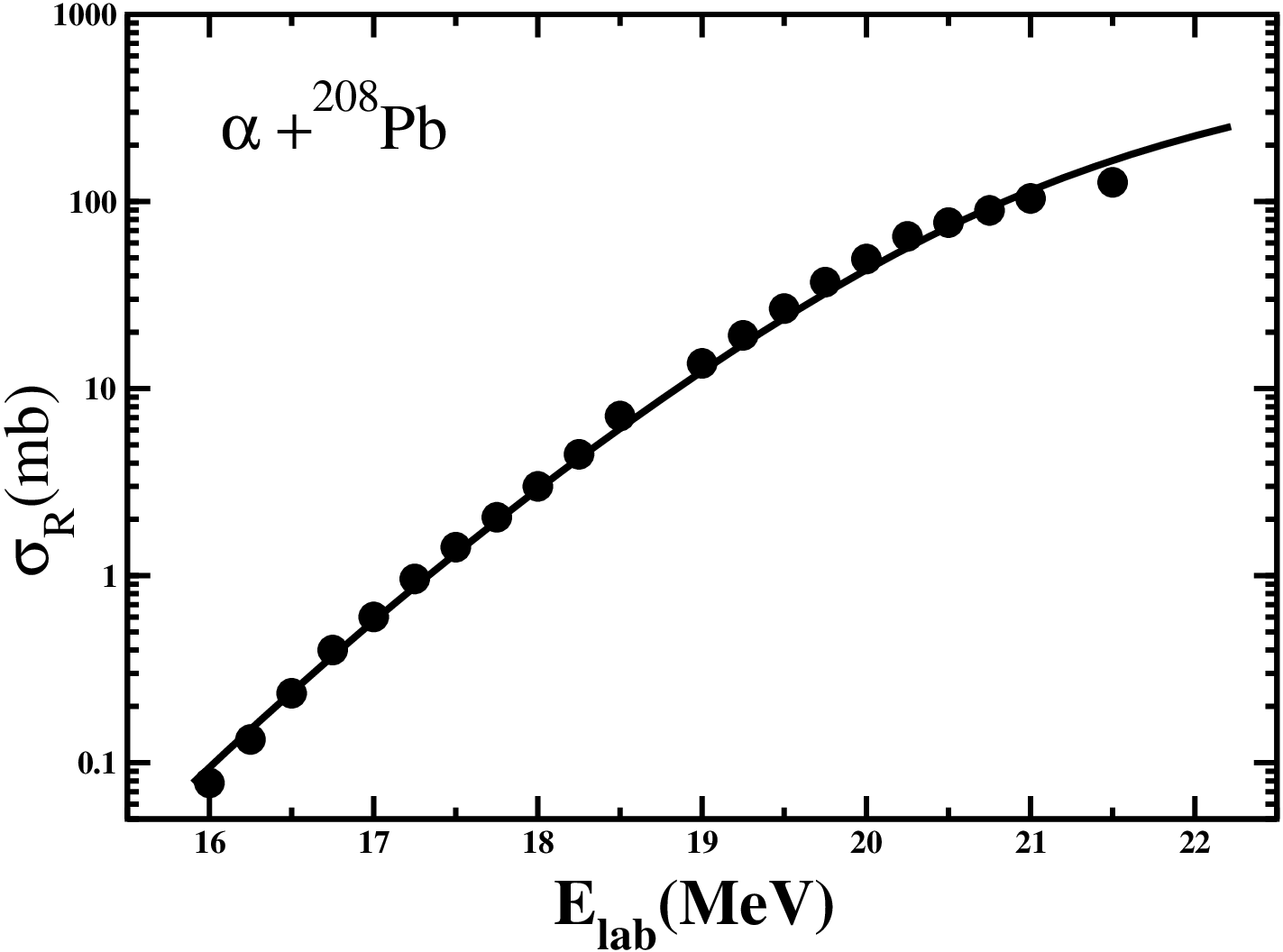}}}
\end{center}
\vspace*{0.2 true in}
\caption{Plot of reaction cross section $\sigma_R$ as function of laboratory
energy for the $\a$+$^{208}$Pb collision. The solid curve represents the 
results of present optical model calculation. The experimental data
shown by solid circles are taken from \cite{bar1974}.
}
\end{figure}

\end{document}